\documentclass{jfm}
\usepackage{graphicx}
\usepackage{epstopdf, epsfig}

\usepackage{amsmath}

\shorttitle{Exact energy stability of B\'enard--Marangoni convection at infinite Prandtl number}
\shortauthor{G. Fantuzzi and A. Wynn}

\title{Exact energy stability of B\'enard--Marangoni convection at infinite Prandtl number}

\author{Giovanni Fantuzzi\aff{1}
  \corresp{\email{gf910@ic.ac.uk}}
  \and
  Andrew Wynn\aff{1}}

\affiliation{\aff{1}Department of Aeronautics, Imperial College London, South Kensington Campus, London SW7 2AZ, UK}

\usepackage{multirow}
\usepackage{pbox}

\newcommand{\BM}{B\'enard--Marangoni}
\newcommand\Ma{\mbox{\textit{M}}}  
\newcommand\B{\mbox{\textit{B}}}  
\renewcommand\L{\mbox{\textit{L}}}  

\newcommand{\MaLi}{\Ma_{\! l}}
\newcommand{\MaEn}{\Ma_{\! e}}

\newcommand{\dz}{\,\mathrm{d}z}
\newcommand{\dt}{\,\mathrm{d}t}

\newcommand{\abs}[1]{\left\vert #1 \right\vert}
\newcommand{\ie}{{\it i.e.}}

\renewcommand{\vec}[1]{\boldsymbol{#1}}

\begin{document}

\maketitle

\begin{abstract}
Using the energy method we investigate the stability of pure conduction in Pearson's model for {\BM} convection in a layer of fluid at infinite Prandtl number. Upon extending the space of admissible perturbations to the conductive state, we find an exact solution to the energy stability variational problem for a range of thermal boundary conditions describing perfectly conducting, imperfectly conducting, and insulating boundaries. 
Our analysis extends and improves previous results, 
and shows that with the energy method global stability can be proven up to the linear instability threshold only when the top and bottom boundaries of the fluid layer are insulating. Contrary to the well-known Rayleigh--B\'enard convection setup, therefore, energy stability theory does not exclude the possibility of subcritical instabilities against finite-amplitude perturbations. 
\end{abstract}


\section{Introduction}
\label{s:intro}

{\BM} convection describes the motion of a layer of fluid driven by shear stresses due to gradients in surface tension at the interface between the fluid and its surroundings. This type of convection arises in numerous engineering applications, including the growth of crystals in semiconductors~\citep{Schatz2001}, cladding processes~\citep{Kumar2009}, and drying of thin polymer films~\citep{Yiantsios2015}, and has recently received increasing attention as a paradigm for shear-driven turbulent transport processes~\citep{Boeck1998,Boeck2001,Hagstrom2010}.

The first mathematical model of surface-tension-driven convection was proposed by~\citet{Pearson1958}, who showed that pure conduction is linearly unstable when the Marangoni number {\Ma}, a non-dimensional measure of the surface tension effects, exceeds a critical threshold $\MaLi$, independently of the fluid's Prandtl number {\Pran} (the ratio of the fluid's kinematic viscosity to its thermal diffusivity). Subsequently,~\citet{Davis1969} used the energy method to prove that conduction is asymptotically stable against disturbances of arbitrary amplitude when $\Ma$ is smaller than a critical value $\MaEn$, also independently of {\Pran}. 
In contrast to Rayleigh--B\'enard convection, the linear and energy thresholds $\MaLi$ and $\MaEn$ do not coincide, allowing the possibility of subcritical instabilities.

Although  Davis's  
analysis and computations yield the best global stability boundary that can be attained with the energy method
for a fluid of finite {\Pran}, they can be improved in the {\it infinite} Prandtl number case. This limit is  an attractive model for high-Prandtl-number fluids, such as the silicon oils used in experiments~\citep{DeBruyn1996} or Earth's mantle~\citep{Jones1977}, because it gives accurate quantitative predictions whilst simplifying the governing equations~\citep{Boeck2001}. 
The key observation is that in the limit of infinite $\Pran$ the inertial term in the momentum equation can be dropped, and as a result the velocity field can be ``slaved"  to the temperature field~\citep[see e.g.][]{Hagstrom2010}, allowing the formulation of an improved variational principle for energy stability. 
This variational principle, first considered by \citet{Hagstrom2010}, is interesting  
from a mathematical perspective 
because it requires the minimisation of a quadratic functional that depends explicitly on the boundary values of the argument function and, as we will show, whose Euler--Lagrange differential equation is overconstrained. 
Hagstrom \& Doering bypassed this difficulty by applying elementary functional estimates to the quadratic functional directly, and raised the energy stability boundary $\MaEn$ from $56.77$ to $58.36$ in the case of a perfectly conducting bottom boundary. 

The main contribution of this work is to show that further improvements are possible.  We consider an extended version of  Hagstrom \& Doering's infinite-{\Pran} energy stability variational principle---which applies to Pearson's model of {\BM} convection with perfectly conducting, imperfectly conducting, and insulating bottom boundaries---and compute the optimal energy stability boundary by extending the domain of the variational problem in such a way that the Euler--Lagrange equation admits a unique solution.

The rest of this work is organised as follows. Section~\ref{s:model} reviews Pearson's model for B\'enard--Marangoni convection at infinite Prandtl number. We formulate the variational principle for energy stability in~\S\ref{e:enStab}, and derive an exact solution in~\S\ref{s:solution}. Further remarks and suggestions for future investigations are offered in~\S\ref{s:conclusion}.

\section{Pearson's model}
\label{s:model}

Consider a layer of fluid of density $\rho$, kinematic viscosity $\nu$, thermal diffusivity $\kappa$, and thermal conductivity $\lambda$, bounded by two non-deformable surfaces at $z=0$ and $z=h$.  When the fluid is at rest and heat is transported by conduction alone, the temperature of the fluid is given by $T(z) = T_0 - (Q_0/\lambda) z$, where $T_0$ is the temperature of the bottom boundary and $Q_0$ is the imposed heat flux through the layer per unit area.
We choose the temperature scale such that $T_0=0$, and make the system non-dimensional by using $h$, $h^2/\kappa$, and $Q_0h/\lambda$ as the characteristic length, time, and temperature units. For simplicity, we work in two dimensions and denote the non-dimensional position vector by $\vec{x}=x\vec{i}+z\vec{k}$; the model and all our results extend with no modifications to three dimensions as described by~\citet{Hagstrom2010}.

At infinite Prandtl number, non-dimensional velocity, pressure, and temperature disturbances to the conductive state---denoted by $\vec{u}(\vec{x},t)=u(\vec{x},t)\vec{i}+w(\vec{x},t)\vec{k}$, $p(\vec{x},t)$, and $\theta(\vec{x},t)$---evolve according to
\begin{subequations}
\begin{align}
\label{e:NSmomentum}
\nabla^2\vec{u} &= \bnabla p,
\\
\label{e:continuity}
\bnabla \bcdot \vec{u} &= 0,
\\
\label{e:pertEq}
\frac{\partial \theta}{\partial t} 
+ \vec{u}\bcdot\bnabla \theta &= \nabla^2 \theta + w.
\end{align}
\end{subequations}

We assume that all variables are periodic in the horizontal ($x$) direction with period $\Lambda$, or that their Fourier transform exists. We impose the no-slip condition $\vec{u}\vert_{z=0} = 0$ at the bottom boundary, and the impenetrability condition $w\vert_{z=1}=0$ at the top boundary.  Moreover, if $\gamma$ denotes the negative of the derivative of surface tension with respect to surface temperature, the balance of surface stresses and tension forces is expressed by the boundary condition
\begin{equation}
\left[ \frac{\partial u}{\partial z} + \Ma\,\frac{\partial \theta}{\partial x}\right]_{z=1}=0,
\end{equation}
where the Marangoni number $\Ma=\gamma Q_0 h^2/(\lambda \rho \nu \kappa)$  is the main governing parameter of the system. Finally, letting $q_{\rm bot}$ and $q_{\rm top}$ denote the derivative of the outward heat fluxes through the top and bottom surfaces with respect to the surface temperature, balancing the heat fluxes through boundaries requires that
\begin{align}
\label{e:BC1}
\left[\frac{\partial \theta}{\partial z} - \B\,\theta\right]_{z=0} &= 0, &
\left[\frac{\partial \theta}{\partial z} + \L\,\theta\right]_{z=1} &= 0,
\end{align}
where the Biot numbers $\B=q_{\rm bot}h/\lambda$, $\L=q_{\rm top}h/\lambda$ describe the conductivity of the boundaries. (The sign difference between the two boundaries is due to the convention that outward heat flux is positive). We consider $\B,\L\geq 0$, a reasonable assumption because an increase in the fluid's surface temperature should raise the heat flux to the surroundings; the case $B=L=0$ corresponds to perfectly insulating boundaries, while the perfectly conducting case corresponds to the (formal) choice $\B=\L=\infty$.  
For a comprehensive discussion of the thermal boundary conditions~\eqref{e:BC1} we refer the reader to the original work by~\citet{Pearson1958}.

\section{Energy stability analysis}
\label{e:enStab}
 
Stability analysis via the energy method relies on the simple observation that stationary conduction ({\ie} when the fluid is at rest) is stable if the kinetic energy of a temperature perturbation does not increase in time, irrespective of the perturbation's initial amplitude. The evolution equation for the average kinetic energy $\langle \theta^2 \rangle/2$ of a temperature perturbation, where $\langle \cdot \rangle$ denotes the usual volume average, is found by averaging $\theta\times$\eqref{e:pertEq} and integrating by parts using~\eqref{e:continuity} and the boundary conditions~\eqref{e:BC1} to arrive at
%
%
\begin{equation}
\label{e:enEq}
\frac{1}{2}\,\frac{{\rm d}}{\dt}\langle \theta^2 \rangle = 
- \langle \abs{\bnabla\theta}^2 - w\theta \rangle 
- \L\,\overline{\theta^2}(1) - \B\,\overline{\theta^2}(0).
\end{equation}
In this equation and throughout the rest of this section, overlines denote horizontal averages.
Clearly, the kinetic energy of the perturbation $\theta$ does not increase in time if the right-hand side of~\eqref{e:enEq} is non-positive at each instant in time, {\ie},
\begin{equation}
\label{e:Cond1}
\langle \abs{\bnabla\theta}^2 - w\theta \rangle 
+ \L\,\overline{\theta^2}(1) + \B\,\overline{\theta^2}(0) \geq 0.
\end{equation}

Upon substituting the horizontal Fourier series expansion of $\theta$ and $w$ into~\eqref{e:Cond1} and into the boundary conditions~\eqref{e:BC1} (we consider the case of a finite periodic domain for definitess; similar arguments hold for the infinite domain if the Fourier series is replaced by the Fourier transform), dropping the time dependence, and recalling from~\cite{Hagstrom2010} that the Fourier amplitudes of the velocity perturbation $w$ are ``slaved'' to those of $\theta$ according to $\hat{w}_k(z)=-\Ma\,f_k(z)\,\hat{\theta}_k(1)$, where
\begin{equation}
\label{e:fk}
f_k(z) = \frac{k\sinh k}{\sinh(2k)-2k} 
\left[ kz\cosh(kz) - \sinh(kz) 
+ (1-k\coth k)\,z \sinh(kz)\right],
\end{equation}
we can rewrite
\begin{equation}
\langle \abs{\bnabla\theta}^2 - w\theta \rangle 
+ \L\,\overline{\theta^2}(1) + \B\,\overline{\theta^2}(0)
= 
2\, \sum_{k \geq 0} \mathcal{F}_k\{\hat{\theta}_k\},
\end{equation}
where the sum is over all positive wavenumbers and
\begin{multline}
\label{e:QkDefOrig}
\mathcal{F}_k\{\hat{\theta}_k\} := \int_0^1 \left\{ 
\vert\hat{\theta}_k'(z)\vert^2 
+ k^2\,\vert\hat{\theta}_k(z)\vert^2 
+\Ma\,f_k(z)\,\Real\left[ {\hat{\theta}_k}^*(z)\,\hat{\theta}_k(1)\right] 
\right\}\!\dz 
\\
+\L\,\vert\hat{\theta}_k(1)\vert^2
+\B\,\vert\hat{\theta}_k(0)\vert^2.
\end{multline}
(Here and in the following, $^*$ denotes complex conjugation and primes denote total differentiation with respect to $z$.)

Since among all possible perturbations are those defined by a single wavenumbers, we conclude that a necessary and sufficient condition for the global stability of {\BM} conduction is that, for all wavenumbers $k\geq 0$,
%
%
\begin{equation}
\label{e:enStabIneq1}
\mathcal{F}_k\{\hat{\theta}_k\} 
\geq 0
\end{equation}
for all complex-valued perturbation Fourier amplitudes $\hat{\theta}_k$ (hereafter simply referred to as perturbations) that satisfy
\begin{align}
\label{e:FTBC}
{\hat{\theta}_k}'(0) - \B\,\hat{\theta}_k(0) &=0, &
{\hat{\theta}_k}'(1) + \L\,\hat{\theta}_k(1) &=0.
\end{align}

In fact, we may restrict our attention to real-valued $\hat{\theta}_k$ because the real and imaginary parts give identical and independent contributions to the left-hand side of~\eqref{e:enStabIneq1}.
Moreover, note that~\eqref{e:enStabIneq1} holds trivially when $\hat{\theta}_k(1)=0$, and that its left-hand side is homogeneous quadratic in $\hat{\theta}_k$. Since the boundary conditions in~\eqref{e:FTBC} are also homogeneous, we can further restrict our attention to the perturbations that satisfy the normalisation condition $\hat{\theta}_k(1)=1$ and, without any loss of generality, we may replace~\eqref{e:BC1} with the boundary conditions
\begin{align}
{\hat{\theta}_k}'(0) - \B\,\hat{\theta}_k(0) &=0,  &
{\hat{\theta}_k}'(1) &= -\L, &
{\hat{\theta}_k}(1) &=1.
\end{align}
Putting these observations together, we define the space of admissible perturbations as
\begin{equation}
\Gamma_0 := \left\{ v(z):\,
\int_0^1\!\left(\vert v' \vert^2+\vert v \vert^2\right)\!\dz<\infty,\,\,
v'(0)=\B\,v(0),\,\,
v'(1)=-\L,\,\,
v(1)=1 \right\}.
\end{equation}

Finally, it is clear that~\eqref{e:enStabIneq1} holds if and only if the \emph{infimum} of its left-hand side over all admissible perturbation fields is non-negative. The stability of {\BM} conduction at given Marangoni and Biot numbers $\Ma$, $\B$, and $\L$ is then established if we can prove that, for all wavenumbers $k$,
%
\begin{equation}
\label{e:QkDef}
\mathcal{Q}_k^\star := \inf_{v\in \Gamma_0}
\int_0^1 \left[ 
\vert v'(z)\vert^2 
+ k^2\,\vert v(z) \vert^2 
+ \Ma\,f_k(z)\, v(z)\right]\!\dz 
+ \L + \B\,\vert v(0)\vert^2
\geq 0.
\end{equation}
In particular, for fixed values of the Biot numbers $\B$ and $\L$ we can compute the energy stability boundary in the {\Ma}--$k$ space---{\ie}, the largest Marangoni number for which a perturbation of wavenumber $k$ is stable---by solving the variational problem for the infimum $\mathcal{Q}_k^\star$ as a function of the Marangoni number $\Ma$ for each $k$ and choosing the largest $\Ma$ for which $\mathcal{Q}_k^\star\geq 0$ .

\section{Solution of the variational problem}
\label{s:solution}

As we have anticipated in \S\ref{s:intro}, the variational problem for $\mathcal{Q}_k^\star$ is interesting from the mathematical point of view because the infimum of the quadratic form
%
\begin{equation}
\label{e:QkDef2}
\mathcal{Q}_k\{v\} = \int_0^1 \left[ \vert v'(z)\vert^2 
+ k^2\,\vert v(z) \vert^2 
+ \Ma\,f_k(z)\, v(z)\right]\!\dz 
+ \L + \B\,\vert v(0)\vert^2
\end{equation}
is not generally attained by any test function $v\in\Gamma_0$. In fact, a straightforward application of the calculus of variations~\citep[see e.g.][]{Courant1953,Giaquinta2004} shows that any candidate minimiser $v_\star$ must satisfy the second-order  Euler--Lagrange differential equation
\begin{equation}
\label{e:EL}
v_\star''(z) - k^2\,v_\star(z) = \frac{1}{2}\,\Ma\,f_k(z),
\end{equation}
subject to the three boundary conditions $v_\star'(0)=\B v_\star(0)$, $v_\star'(1)=-\L$, and $v_\star(1)=1$. This problem is over-constrained, and admits no solution (with the possible exception of selected values of $\B$, $\L$, $\Ma$ and $k$). 

It should be noted that the lack of a minimiser for $\mathcal{Q}_k$ is not due to our normalisation convention for the test functions, which is the source of the extra boundary condition $v_\star(1)=1$. When a different normalisation is used, in fact, the minimisation of $\mathcal{Q}_k$ over $\Gamma_0$ is replaced with the minimisation of $\mathcal{F}_k$ in~\eqref{e:QkDefOrig} over all normalised test functions that satisfy the boundary conditions~\eqref{e:FTBC}. As we demonstrate in appendix~\ref{a:IllPosed} for the commonly used normalisation $\int_0^1 \vert v(z)\vert^2 \dz = 1$, the Euler--Lagrange equations for the minimiser of $\mathcal{F}_k$ are over-constrained by so-called ``natural conditions" that arise when setting to zero its first variation~\citep[][Chapter IV, Section 5.1]{Courant1953}. 

This obstacle is overcome if we can enlarge the space of test functions in such a way that~\eqref{e:EL} has a unique solution $v_\star$, and moreover $\mathcal{Q}_k\{v_\star\}=\mathcal{Q}_k^\star$. This is indeed the case if we drop the boundary condition $v'(1)=-\L$, and minimise $\mathcal{Q}_k\{v\}$ over the {\it larger} space of functions
\begin{equation}
\Gamma_1 := \left\{ 
v(z) :\,
\int_0^1\!\left( \vert v' \vert^2+\vert v \vert^2 \right)\dz<\infty,\,\,
v'(0)=\B\,v(0),\,\,
v(1)=1 
\right\}.
\end{equation}
Having removed one boundary condition, in fact,  the Euler--Lagrange equation~\eqref{e:EL} becomes a standard second-order inhomogeneous ordinary differential equation and it can be solved analytically. For each wavenumber $k$ the solution can be written in the form
\begin{equation}
\label{e:vstar}
v_\star(z) = \Ma\, g_k(z) + h_k(z),
\end{equation}
where $g_k(z)$ and $h_k(z)$ are two known smooth functions whose expressions, given in Appendix~\ref{a:ghexpr}, depend on the Biot number $\B$.
Furthermore, to see that $\mathcal{Q}_k^\star = \mathcal{Q}_k\{v_\star\}$ we note that on one hand we must have $\mathcal{Q}_k\{v_\star\}\leq \mathcal{Q}_k^\star$, because $\Gamma_0$ is a proper subset of $\Gamma_1$ and $v_\star$ minimises $\mathcal{Q}_k\{v\}$ over $\Gamma_1$. On the other hand, $\mathcal{Q}_k\{v_\star\} \geq Q_k^\star$ because we can find a sequence of functions $(v_n)_{n\geq 1}$ with $v_n\in\Gamma_0$ such that $\mathcal{Q}_k\{v_n\}$ converges to $\mathcal{Q}_k\{v_\star\}$; for example, in Appendix~\ref{a:ConvergenceProof} we show that this is the case if we let $\xi_n = n/(n+1)$ and take
\begin{equation}
\label{e:vn}
v_n(z) := \begin{cases}
v_\star(z) &\text{if}\quad\displaystyle 0\leq z \leq \xi_n,
\\\displaystyle
v_\star(\xi_n) + 
\frac{2+\L\,(1-\xi_n) - 2\,v_\star(\xi_n)}{1-\xi_n}\left(z - \xi_n\right)
&\text{if}\quad\displaystyle \xi_n \leq z \leq \frac{1+\xi_n}{2},
\\\displaystyle
1+\L\,(1-z), &\text{if}\quad\displaystyle\frac{1+\xi_n}{2} \leq z \leq 1.
\end{cases}
\end{equation}
Note that these test functions are simply piecewise-linear continuous functions on $[\xi_n,1]$ satisfying $v_n(\xi_n)=v_\star(\xi_n)$, $v_n(1)=1$, and $v_n'(1)=-L$, and that they could be smoothed around the corner points without changing their boundary values and derivatives to meet any regularity requirements prescribed on the space $\Gamma_0$. 

Having computed the minimizer $v_\star$, we now turn to the computation of the minimum $\mathcal{Q}_k^\star=\mathcal{Q}_k\{v_\star\}$. To simplify the analysis, we integrate $v_\star\times$\eqref{e:EL} by parts using the boundary conditions on $\Gamma_1$ to show that
%
\begin{equation}
\int_0^1 \left[
\vert v_\star'(z)\vert^2 + k^2 \vert v_\star(z) \vert^2
\right]\!\dz  = 
v_\star'(1) - \B\,\vert v_\star(0)\vert^2 - \frac{1}{2}\,\Ma \int_0^1 f_k(z)\, v_\star(z) \dz.
\end{equation}
Upon combining this with~\eqref{e:QkDef2} and~\eqref{e:vstar} we find
\begin{equation}
\label{e:Qstar}
\mathcal{Q}_k^\star  = 
\underbrace{\left[ \frac{1}{2}\int_0^1 g_k(z)\,f_k(z)\dz \right]}_{=:\alpha_k} \Ma^2
+ \underbrace{\left[ {g_k}'(1) + \frac{1}{2}\int_0^1 h_k(z)\,f_k(z)\dz\right]}_{=:\beta_k} \Ma
+ {h_k}'(1) + \L.
\end{equation}

\begin{figure}
\centering
\includegraphics[scale=1.05]{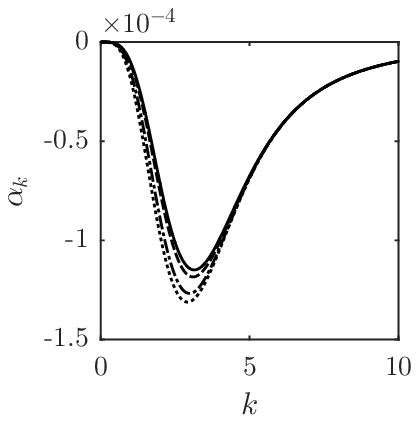}\hfill
\includegraphics[scale=1]{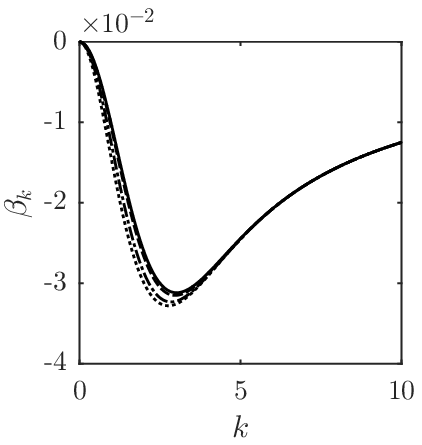}\hfill
\includegraphics[scale=1]{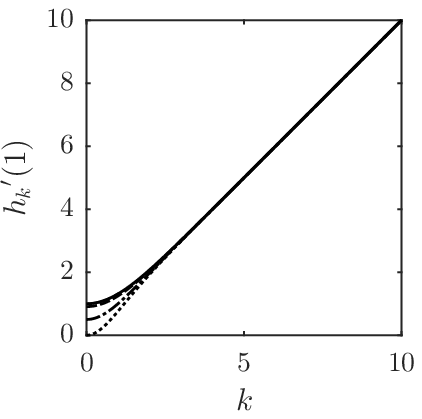}
\caption{Value of the coefficients $\alpha_k$ and $\beta_k$, defined as in~\eqref{e:Qstar}, and of $h_k'(1)$, plotted as a function of the wavenumber $k$ for $\B = 0$ (dotted line), $\B=1$ (dot-dashed line), $\B=10$ (dashed line) and $\B = \infty$ (solid line).}
\label{f:coeffs}
\end{figure}

For each wavenumber $k$ and given Biot numbers $\B$ and $\L$, the infimum $\mathcal{Q}_k^\star$ is a quadratic form of the Marangoni number $\Ma$, and the coefficients $\alpha_k$ and $\beta_k$ have explicit expressions ($f_k$, $g_k$ and $h_k$ are known, and their products can be integrated analytically). These are too long to be reported, but $\alpha_k$ and $\beta_k$, together with $h_k'(1)$, are plotted in figure~\ref{f:coeffs} for Biot numbers $\B=0$ (corresponding to a perfectly insulating bottom boundary), $\B=1$, $\B=10$, and $\B=\infty$ (corresponding to a perfectly conducting bottom boundary).  Note that the leading order  coefficient $\alpha_k$ is negative for all $k$, and it must be so because perturbations at any wavenumber $k$ eventually become linearly unstable~\citep{Pearson1958}, implying that $\mathcal{Q}_k^\star < 0$ for all  sufficiently large $\Ma$. Consequently, the largest Marangoni number $\MaEn$ at which {\BM} conduction at infinite Prandtl number is stable against perturbations of wavenumber $k$ and arbitrary amplitude is given by the largest root of the quadratic form in~\eqref{e:Qstar}, {\ie}
\begin{equation}
\MaEn = \frac{-\beta_k - \sqrt{\beta_k^2 - 4\,\alpha_k\left[h_k'(1) + \L\right]} }{2\,\alpha_k}.
\end{equation}

\begin{figure}
\centering
\includegraphics[scale=1]{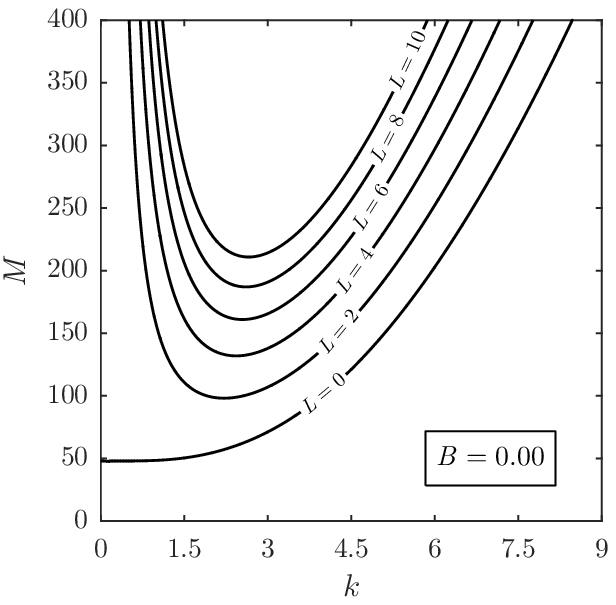}\hspace{1em}
\includegraphics[scale=1]{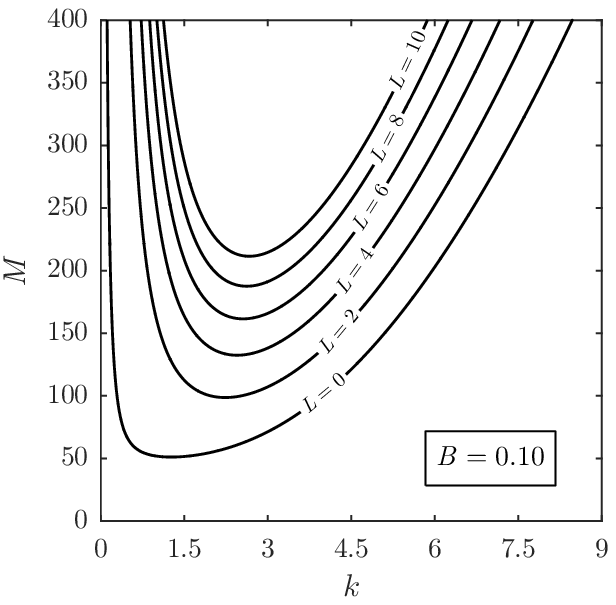}\\[1em]
\includegraphics[scale=1]{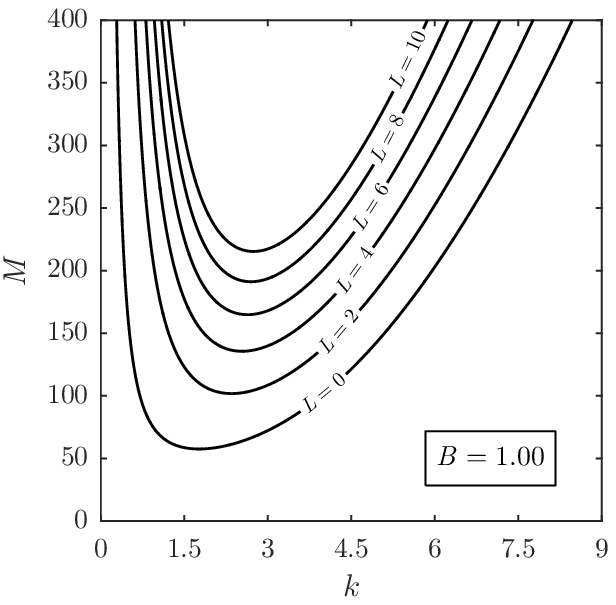}\hspace{1em}
\includegraphics[scale=1]{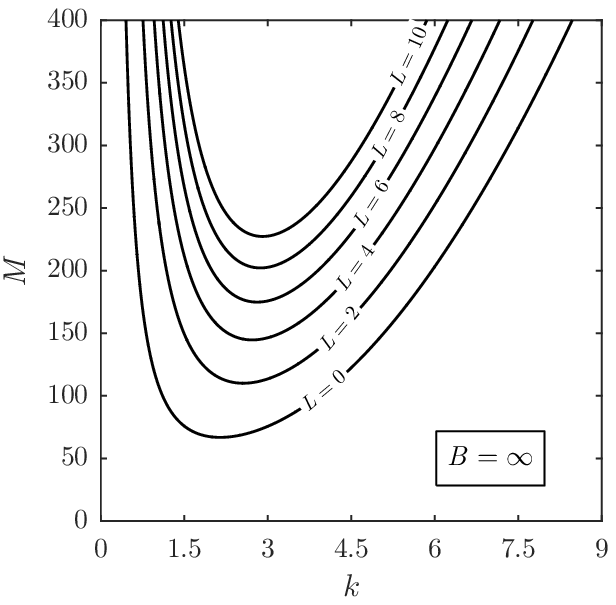}
\caption{Critical energy stability curves for {\BM} conduction at infinite Prandtl number in the {\Ma}--$k$ space, computed using equation~\eqref{e:Qstar} for four different values of the Biot number of the bottom surface, $\B$, and six values of the Biot number of the top surface, $\L$. The extremal cases $\B=0$ and  $\B=\infty$ correspond to a perfectly insulating and a perfectly conducting bottom boundary, respectively.}
\label{f:results}
\end{figure}

Figure~\ref{f:results} illustrates the optimal stability boundary in the {\Ma}--$k$ space, given by the curve $\MaEn(k)$, for fixed values of the Biot numbers $\B$ and $\L$. As in the linear stability analysis of~\citet{Pearson1958}, increasing the Biot number $\L$ of the upper surface raises the critical Marangoni number. This fact is obvious from~\eqref{e:Qstar}, and it corresponds to the physical observation that improving the conductivity of the upper boundary reduces the surface temperature gradients and, consequently, the surface tension driving the flow~\citep{Davis1987}. 
Also analogous to the linear stability problem is the fact that in the case of two insulating boundaries ($\B=\L=0$) the minimum critical Marangoni number $\MaEn =  48$ is achieved for $k=0$. Interestingly, this coincides with Pearson's linear stability threshold~\citep{Pearson1958},  {\ie}, {\BM} conduction between insulating boundaries is globally stable until an ``infinite wavelength'' linear instability occurs. This instability is suppressed by any increase in $\B$, and the qualitative distribution of the energy stability boundaries for an imperfectly conducting bottom boundary ($\B$ finite) is the same as in the perfectly conducting case ($\B=\infty$).

Table~\ref{t:results}  presents the minimum critical Marangoni number over all wavenumbers $k$, denoted $\Ma_{\rm cr}$, and the critical wavenumber $k_{\rm cr}$ for selected Biot numbers $\L$ in the extremal cases $\B=0$ (insulating bottom boundary) and $\B=\infty$ (perfectly conducting bottom boundary). These values are compared to the corresponding linear stability results from~\citet{Pearson1958} and, when available, to the energy stability results obtained by~\citet{Davis1969} for finite-{\Pran} fluids: since these are actually independent of the Prandtl number, they also apply in the infinite-{\Pran} case. As one would expect, our values are larger than those computed by Davis, because the infinite-$\Pran$ variational problem exploits the explicit coupling between the velocity and temperature fields. Moreover, for $\B=\infty$ and $\L=0$ we find $\Ma_{\rm cr}=66.84$, a 14.5\% improvement on the value $58.36$ computed by~\citet{Hagstrom2010}. On the other hand,  the optimal energy stability boundary is strictly smaller than the linear stability one, with the only exception of the case $\B=\L=0$ (note that Pearson's linear stability analysis is unchanged when $\Pran=\infty$). This means that, unlike in Rayleigh--B\'enard convection, there generally exists a finite range of Marangoni numbers for which the flow is linearly stable, but subcritical instabilities due to perturbations of finite amplitude may occur.

\begin{table}
  \begin{center}
  \begin{tabular}{rrrrrrrrrrrr}
  			  &$\phantom{00}$&&$\phantom{00}$&
  			  \multicolumn{2}{c}{
  			  \pbox{4cm}{\relax\ifvmode\centering\fi
  			  Energy stability \\ at $\Pran=\infty$}}
  			  &$\phantom{00}$&
  			  \multicolumn{2}{c}{\pbox{4cm}{\relax\ifvmode\centering\fi
  			  Linear stability\\ \citep{Pearson1958}}}
  			  &$\phantom{00}$&
  			  \multicolumn{2}{c}{\pbox{4cm}{\relax\ifvmode\centering\fi
  			  Energy stability\\ \citep{Davis1969}}}\\[1em]
  			  \cline{5-6} \cline{8-9} \cline{11-12}\\[-0.5em]
  			  $\B$&&$\L$&& $\Ma_{\rm cr}$ &  $k_{\rm cr}$ &
  			  & $\Ma_{\rm cr}$ &  $k_{\rm cr}$ &
  			  & $\Ma_{\rm cr}$ &  $k_{\rm cr}$ \\[1em]
               0 && 0 & & 48 & 0 & & 48 & 0 & & --- & --- \\
               0 && 10 & & 210.9 & 2.65 & & 383.2 & 2.45 && --- &--- \\
               0 && 100 & & 769.8 & 2.87 && 3132 & 2.71 && --- & ---\\[1em]
     $\infty$&& 0 && 66.84 & 2.14 && 79.61 & 1.99 && 56.77 & ---\\
     $\infty$&& 10 && 227.3 & 2.91 && 413.4 & 2.74 && 180.7 & ---\\
     $\infty$&& 100 && 821.6 & 2.98 && 3304 & 3.10 && --- & ---\\[0.5em]
  \end{tabular}
  \caption{Minimum critical Marangoni number for energy stability at infinite Prandtl number for selected values of the Biot number $\L$ in the extremal cases $\B=0$ (insulating bottom boundary) and $\B=\infty$ (perfectly conducting bottom boundary). Where available, the corresponding values for linear stability~\citep{Pearson1958} and energy stability at finite $\Pran$~\citep{Davis1969} are also reported.}
\label{t:results}
\end{center}
\end{table}

\section{Conclusion}
\label{s:conclusion}

To summarise, we have studied the global stability of the purely conductive state of infinite-Prandtl-number {\BM} convection using the method of energy, and we have computed the exact critical Marangoni number in wavenumber space for thermal boundary conditions corresponding to perfectly conducting,  imperfectly conducting, and perfectly insulating boundaries. We have shown that in the infinite-$\Pran$ limit, the explicit slaving of the velocity field to the temperature field can be exploited to raise the energy stability boundary compared to the finite {\Pran} case, although a gap with the linear stability threshold remains in all but the insulating case $\B=\L=0$. Whether global stability attains up to the linear stability boundary or subcritical instabilities exist, should be determined by bifurcation analysis, numerical simulations, or alternative techniques for global stability analysis, such as those of~\citet{Goulart2012} and~\citet{Chernyshenko2013}.

Finally, we note that the analysis presented in this work may be of use in the computation of upper bounds on the convective heat transport using the background field method \citep[see e.g.][]{Constantin1995,Constantin1995a,Doering1992,Doering1994,Doering1996}. The method, already applied to {\BM} convection by~\citet{Hagstrom2010}, relies on the construction of a background temperature field $\tau(z)$, subject to a nonlinear stability condition obtained by replacing $\Ma\,f_k(z)$ with $2\,\Ma\,\tau'(z)\,f_k(z)$ in the energy stability constraint~\eqref{e:QkDef}. Given a candidate background field $\tau$, this nonlinear condition can be analysed using the same ideas presented in \S\ref{s:solution}, and the corresponding the Euler--Lagrange equation has an analytic solution. Whether this allows one to lower Hagstrom \& Doering's original bound, in the same way that their energy stability result was improved in this work, remains an intriguing open question for future work.

\appendix

\section{On the issue of ``natural conditions"}
\label{a:IllPosed}

Upon restricting attention to real-valued perturbations,~\eqref{e:enStabIneq1} implies that the conduction solution is stable if the functional $\mathcal{F}_k$ in~\eqref{e:QkDefOrig} satisfies
\begin{equation}
\label{e:F}
\inf_{v} \mathcal{F}_k\{v\} \geq 0,
\end{equation} 
the infimum being taken over all square-integrable test functions $v$ with a square-integrable (weak) first derivative, and that satisfy the boundary conditions in~\eqref{e:FTBC}.
Instead of the normalisation condition $v(1)=1$ assumed in~\S\S\ref{s:model}-\ref{s:solution}, suppose we normalise $v$ such that
\begin{equation}
\label{e:normCond}
\int_0^1 \vert v(z)\vert^2 \dz = 1.
\end{equation}
This choice of normalisation is legitimate because the problem is homogeneous, and because $\mathcal{F}_k\{0\}=0$. Letting $\lambda$ be the Lagrange multiplier enforcing~\eqref{e:normCond}, the Euler--Lagrange equations for a candidate minimiser of $\mathcal{F}_k$ are found by setting to zero the first variations of the augmented functional
\begin{equation}
\mathcal{L}_k\{v,\lambda\} :=
\mathcal{F}_k\{v\} + \lambda \left( \int_0^1 \vert v(z)\vert^2 \dz- 1\right).
\end{equation}
Setting to zero the first variation of $\mathcal{L}_k$ with respect to $\lambda$ simply yields~\eqref{e:normCond}. Moreover, the necessary condition for $\mathcal{L}_k$ to be stationary with respect to $v$ is that
\begin{equation}
\frac{\delta \mathcal{L}_k}{\delta v} 
:= \lim_{\varepsilon\to 0} 
\frac{\mathcal{L}_k\{v+\varepsilon\,h,\lambda\} - \mathcal{L}_k\{v,\lambda\}}{\varepsilon} = 0
\end{equation}
for any function $h(z)$ that satisfies $h'(0) - \B\,h(0)=0$ and $h'(1)+\L\,h(1)=0$. Upon integrating by parts using the boundary conditions on $v$, we find
\begin{multline}
\frac{\delta \mathcal{L}_k}{\delta v} 
= \int_0^1 \left[ 
-2\,v''(z) 
+ 2\left(k^2+\lambda\right)\,v(z)
+\Ma\,f_k(z)\,v(1)
\right] h(z)\dz \\
+\left[ \L\,v(1) + \Ma\,\int_0^1 f_k(z)\,v(z)\,dz \right]\,h(1)
+\B\,v(0)\,h(0).
\end{multline}
After requiring the right-hand side above to vanish for all perturbations $h$, we conclude that the minimiser of $\mathcal{F}_k$ subject to~\eqref{e:normCond} must satisfy the Euler--Lagrange equation
\begin{equation}
\label{e:ELapp1}
-2\,v''(z) + 2\left(k^2+\lambda\right)\,v(z) +\Ma\,f_k(z)\,v(1) = 0,
\end{equation}
as well as the ``natural conditions"
%
\begin{align}
\label{e:ELapp2}
\L\,v(1) + \Ma\,\int_0^1 f_k(z)\,v(z)\,dz &=0,
&
\B\,v(0) &= 0,
\end{align}
%
the normalisation condition~\eqref{e:normCond}, and the original boundary conditions
\begin{align}
\label{e:ELapp4}
v'(0) - \B\,v(0) &=0,&
v'(1) + \L\,v(1) &= 0.
\end{align}
Note that no solution exists in general: there are three equations for the two unknowns $v$ and $\lambda$, and furthermore there are three boundary conditions---the two original ones plus the ``natural'' boundary condition $\B\,v(0) = 0$---but only two integration constants for $v$.

\section{Expressions for $g_k(z)$ and $h_k(z)$}
\label{a:ghexpr}
The expressions for $h_k(z)$ and $g_k(z)$ are
\begin{align}
h_k(z) &= \frac{\B\,\sinh(kz)+k\,\cosh(kz)}{\B\,\sinh k +k\,\cosh k},
\\[0.35em]
g_k(z) &= \frac{ P_1(z)\,\cosh(kz) + P_2(z)\,\sinh(kz)}{8\,k\,\left[ \sinh(2k)-2k \right]\left( \B\,\sinh k +k\,\cosh k \right)},
\end{align}
where
\begin{align}
P_1(z) := &-k\,\sinh k\left[ ( k\,\cosh k - \sinh k ) z +3\,\sinh k \right] z\, \B
\nonumber \\
&-k^2\,\cosh k (k\,\cosh k - \sinh k )z^2
- 3\,k^2\,\cosh k\,\sinh k\,z
\nonumber \\
&+ k\,( k\,\cosh k \,\sinh k + k^2 - 2\,\vert \cosh k\vert^2 + 2),
\end{align}
%
and
\begin{align}
P_2(z) := &\left[ k^2\,\vert\sinh k\vert^2\,z^2 
+ \sinh k\,(k\,\cosh k-\sinh k)\,z 
\right.
\nonumber \\
&\left.
+ k\,\cosh k \,\sinh k + k^2 + \vert\cosh k\vert^2-1
\right]\,\B
+
k^3\,\cosh k\,\sinh k\,z^2
\nonumber \\
&+k\,\cosh k (k\,\cosh k - \sinh k)\,z
+ 3\,k\,\cosh k \,\sinh k.
\end{align}

\section{Convergence of $\mathcal{Q}_k\{v_n\}$ to $\mathcal{Q}_k^\star$}
\label{a:ConvergenceProof}

The test function $v_n$ in~\eqref{e:vn} belongs to the functional space $\Gamma_0$ because it satisfies the boundary conditions prescribed on $\Gamma_0$, it is square integrable on the interval $[0,1]$, and so is its (weak) derivative
\begin{equation}
v_n'(z) := \begin{cases}
v_\star'(z) &\text{if}\quad\displaystyle 0\leq z \leq \xi_n,
\\\displaystyle
\frac{2+\L\,(1-\xi_n) - 2\,v_\star(\xi_n)}{1-\xi_n}
&\text{if}\quad\displaystyle \xi_n \leq z \leq \frac{1+\xi_n}{2},
\\\displaystyle
-L, &\text{if}\quad\displaystyle\frac{1+\xi_n}{2} \leq z \leq 1.
\end{cases}
\end{equation}
The convergence of $\mathcal{Q}_k\{v_n\}$ to $\mathcal{Q}_k^\star$ follows from a relatively straightforward application of Lebesgue's dominated convergence theorem. For example,
note that
$\vert v_n'(z) \vert^2 \to \vert v_\star'(z)\vert^2 $ pointwise in $(0,1)$ as $n\to\infty$ since $\xi_n=n/(n+1)\to 1$, and that there exists a constant $C_0>0$ such that $\vert v_n' \vert^2 \leq C_0$ for all $z\in(0,1)$ and $n\geq 1$ since {\it(i)} $\vert v_\star'(z) \vert \leq C_1$ for all $z\in(0,\xi_n)$ for some constant $C_1>0$ because $v_\star$ is a smooth function, and {\it(ii)} by virtue of Taylor's theorem there exists $\eta\in[\xi_n,1]$ such that
\begin{equation}
\left\vert
\frac{2+\L\,(1-\xi_n) - 2\,v_\star(\xi_n)}{1-\xi_n} 
\right\vert
=
\left\vert
\frac{2+\L\,(1-\xi_n) - 2\,[1 - v_\star'(\eta)(1-\xi_n)]}{1-\xi_n} 
\right\vert
\leq 
\L + 2\,C_1.
\end{equation}
Lebesgue's dominated convergence theorem then implies that $\int_0^1 \vert v_n'\vert^2 \dz \to \int_0^1 \vert v_\star' \vert^2 \dz$ as $n\to\infty$. Similar arguments can be applied to $\vert v_n\vert^2$ and $f_k v_n$.

\bibliographystyle{jfm}
\bibliography{refs}

\end{document}